\documentclass[twocolumn,showpacs,prb]{revtex4}

\usepackage{amssymb}

\usepackage{graphicx,amsmath}

\begin{document}

\title{Split-gate quantum point contacts with tunable channel length}

\author{M.~J.~Iqbal}
\email{javaid2k@gmail.com}
\author{J.~P.~de.~Jong}
\affiliation{Zernike Institute for Advanced Materials, Nijenborgh 4, University of
Groningen, NL-9747AG Groningen, The Netherlands}
\author{D.~Reuter}
\author{A.~D.~Wieck}
\affiliation{Angewandte Festk\"{o}rperphysik, Ruhr-Universit\"{a}t
Bochum, D-44780 Bochum, Germany}
\author{C.~H.~van~der~Wal}
\affiliation{Zernike Institute for Advanced Materials, Nijenborgh 4, University of
Groningen, NL-9747AG Groningen, The Netherlands}

\date{version \today}

\begin{abstract}
We report on developing split-gate quantum point contacts (QPCs)
that have a tunable length for the transport channel. The QPCs were
realized in a GaAs/AlGaAs heterostructure with a two-dimensional
electron gas (2DEG) below its surface. The conventional design uses
2 gate fingers on the wafer surface which deplete the 2DEG
underneath when a negative gate voltage is applied, and this allows
for tuning the width of the QPC channel. Our design has 6 gate
fingers and this provides additional control over the form of the
electrostatic potential that defines the channel. Our study is based
on electrostatic simulations and experiments and the results show
that we developed QPCs where the effective channel length can be
tuned from about 200~nm to 600~nm. Length-tunable QPCs are important
for studies of electron many-body effects because these phenomena
show a nanoscale dependence on the dimensions of the QPC channel.
\end{abstract}


\maketitle




\section{\label{5sec:Intro}Introduction}

\begin{figure}[t!]
\centering
\includegraphics[width=1.00\columnwidth]{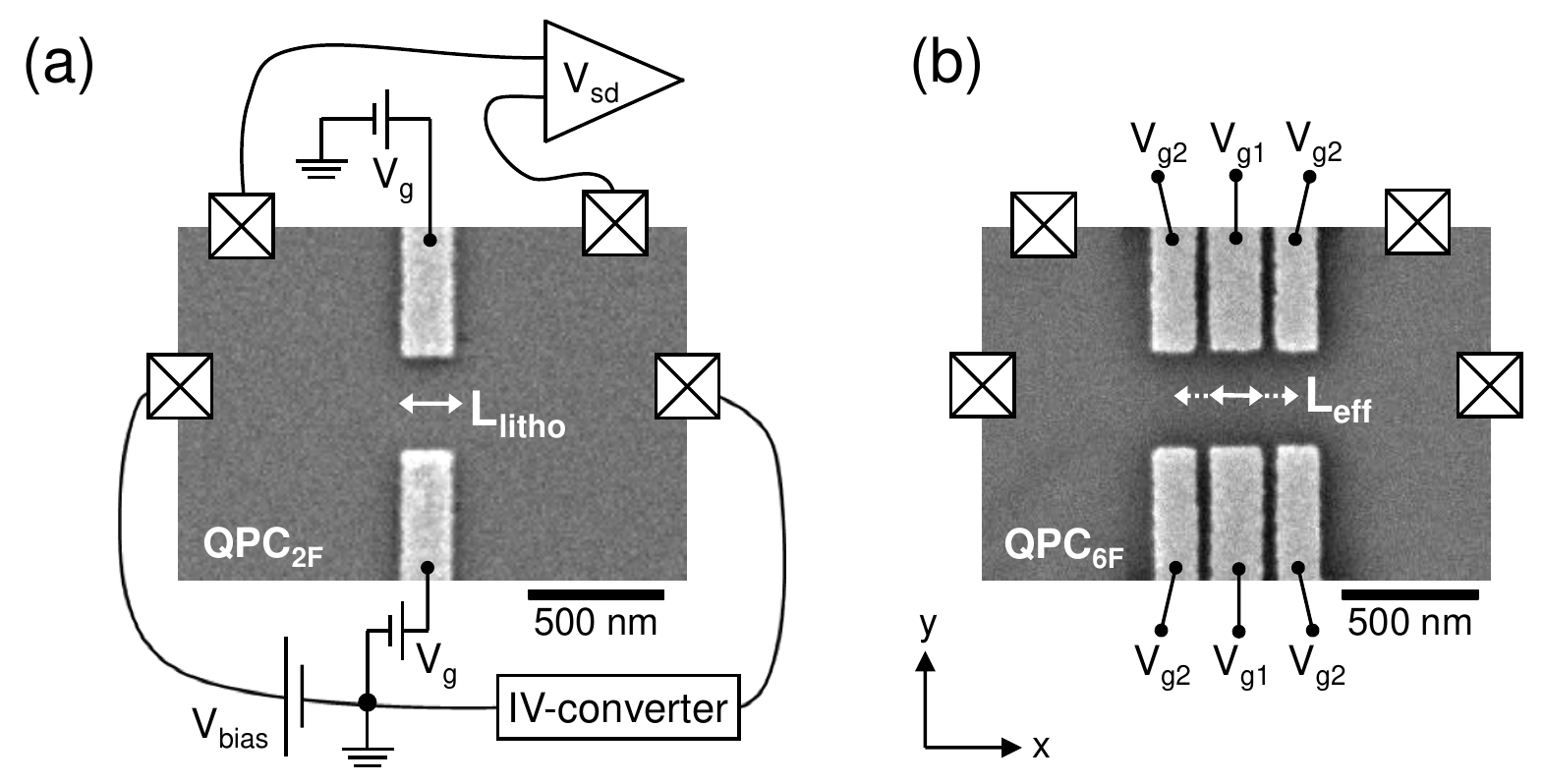}
\caption{(a) SEM image of a conventional split-gate quantum point
contact (QPC). It has two gate fingers (QPC$_{\rm 2F}$ device). The
length of the QPC channel is fixed and can be parameterized by the
lithographic length $L_{litho}$ of the gate structure. The diagram
also illustrates the measurement scheme (see main text for details).
(b) SEM image of a length tunable QPC with 6 gate fingers (QPC$_{\rm
6F}$ device). Here the effective length $L_{eff}$ of the QPC can be
tuned by changing the ratio of the gate voltages on the central
gates ($V_{g1}$) and side gates ($V_{g2}$).} \label{fig1SEMpic}
\end{figure}

A quantum point contact (QPC) is the simplest mesoscopic device that
directly shows quantum mechanical properties. It is a short
ballistic transport channel between two electron reservoirs, which
shows quantized conductance as a function of the width of the
channel \cite{5Wees,5Wharam}.  A widely applied approach for
implementing QPCs is using a split-gate structure on the surface of
a heterostructure with a two-dimensional electron gas (2DEG) at
about 100~nm beneath its surface. The conventional design of such a
split-gate QPC has two metallic gate fingers
(Fig.~\ref{fig1SEMpic}a). Operating this device with a negative gate
voltage $V_g$ results in the formation of a barrier with a small
tunable opening between two 2DEG reservoirs, because the 2DEG below
the gate fingers gets depleted over a range that depends on $V_g$.
For electrons in the 2DEG, this appears as an electrostatic
potential $U$ that is a large barrier with a small opening in the
form of a saddle-point potential (Fig.~\ref{fig3SaddlePotential}).
The saddle-point potential gives transverse confinement in the
channel that is roughly parabolic, which results for this transverse
direction in a discrete set of electronic energy levels. For
electron transport along the channel this gives a discrete set of
subbands with one-dimensional character. Quantized conductance
appears because each subband contributes $G_{0} = 2e^2/h$ to the
channel's conductance \cite{5Wees,5Wharam}, where $e$ is the
electron charge and $h$ is Planck's constant.


We present here the design and experimental characterization of QPCs
which offer additional control over the shape of the saddle-point
potential. We focused on developing devices for which the effective
length of the saddle-point potential (along the transport direction)
can be tuned \textit{in situ}. The additional control is implemented
with a symmetric split-gate design based on 6 gate fingers
(Fig.~\ref{fig1SEMpic}b). Such devices will be denoted as QPC$_{\rm
6F}$, and conventional devices with 2 gate fingers
(Fig.~\ref{fig1SEMpic}a) as QPC$_{\rm 2F}$. These QPC$_{\rm 6F}$ are
operated with the gate voltage on the outer fingers ($V_{g2}$) less
negative than the gate voltage on the central fingers ($V_{g1}$) to
avoid quantum dot formation. Sweeping $V_{g1}$ from more to less
negative values opens the QPC$_{\rm 6F}$. By co-sweeping $V_{g2}$ at
fixed ratio $V_{g2}/V_{g1}$ it behaves as a QPC with a certain
length for the saddle-point potential, and this length can be chosen
by setting $V_{g2}/V_{g1}$: It is shortest for $V_{g2}/V_{g1}
\approx 0$ and longest for $V_{g2}/V_{g1} \lessapprox 1$. For our
design the effective length could be tuned from about 200~nm to
600~nm.

The motivation for developing these length-tunable QPCs comes from
studies of electron many-body effects in QPCs. A well-known
manifestation of these many-body effects is the so-called
0.7~anomaly \cite{5Thomas1996}, which is an additional shoulder at
$0.7~G_{0}$ in quantized conductance traces. These many-body effects
are, despite many experimental and theoretical studies since 1996
\cite{5Review0.7}, not yet fully understood. Recent theoretical work
\cite{5Meir} suggested that many-body effects cause the formation of
one or more self-consistent localized states in the QPC channel, and
that these effects result in the 0.7~anomaly and the other
signatures of many-body physics. This theoretical work predicted a
clear dependence on the length of the QPC channel, and testing this
directly requires experiments where this length is varied.

The work by Koop \textit{et al.} \cite{5Koop} already explored the
relation between the device geometry and parameters that describe
the many-body effects in a large set of QPC$_{\rm 2F}$ devices. This
work compared nominally identical devices, and devices for which the
lithographic length $L_{litho}$ (see Fig.~\ref{fig1SEMpic}a) and
width of the channel in the split-gate structure was varied. These
results were, however, not conclusive. The parameters that describe
the many-body effects showed large, seemingly random variation, not
correlated with the device geometry. At the same time, the devices
showed (besides the 0.7 anomaly) clean quantized conductance traces,
and the parameters that reflect the non-interacting electron physics
did show the variation that one expects when changing the geometry
(for example, the channel pinch-off gate voltage $V_{po}$ and
subband spacing $\hbar \omega_{12}$). This confirms that these QPCs
had saddle-point potentials that were smooth enough for showing
quantized conductance, while it also shows that the many-body
effects are very sensitive to small static fluctuations on these
saddle-point potentials or to nanoscale device-to-device variations
in the dimensions of the potentials. This picture was confirmed by
shifting the channel position inside a particular QPC$_{\rm 2F}$
device. This can be implemented by operating a QPC$_{\rm 2F}$ with a
difference $\Delta V_g$ between the values of $V_g$ on the two gate
fingers in Fig.~\ref{fig1SEMpic}a. Such a channel shift did not
change the quantized conductance significantly, but did cause strong
variation in the signatures of many-body physics. Earlier work had
established that such device-to-device fluctuations can be due to
remote defects or impurities, a slight variation in electron
density, or due to the nanoscale variation in devices that is
inherent to the nanofabrication process
\cite{5Koop,5DefectsInteractions,5JohnAltDavies}. Consequently,
studying how the many-body effects depend on the length of the QPC
channel requires QPCs for which the channel length can be tuned
continuously \textit{in situ}, and where this can be operated
without a transverse displacement of the QPC channel in the
semiconductor material. The work that we report here aimed at
realizing such devices.

This article is organized as follows:
Section~\ref{5sec:designConsids} starts with a short overview of the
options and the choices we made for realizing the QPC$_{\rm 6F}$
devices. Next, in Section~\ref{5sec:Simulations}, we present the
results of electrostatic simulations. In
Section~\ref{5sec:sampleFabAndTech}, we describe the sample
fabrication and measurement techniques. This is followed by
comparing results from simulations and experiments for QPC$_{\rm
6F}$ devices in Section \ref{5sec:ExperimentalRelization}, and
Section~\ref{5sec:Conclusion} summarizes our conclusions.


\begin{figure}[t!]
\centering
\includegraphics[width=0.85\columnwidth]{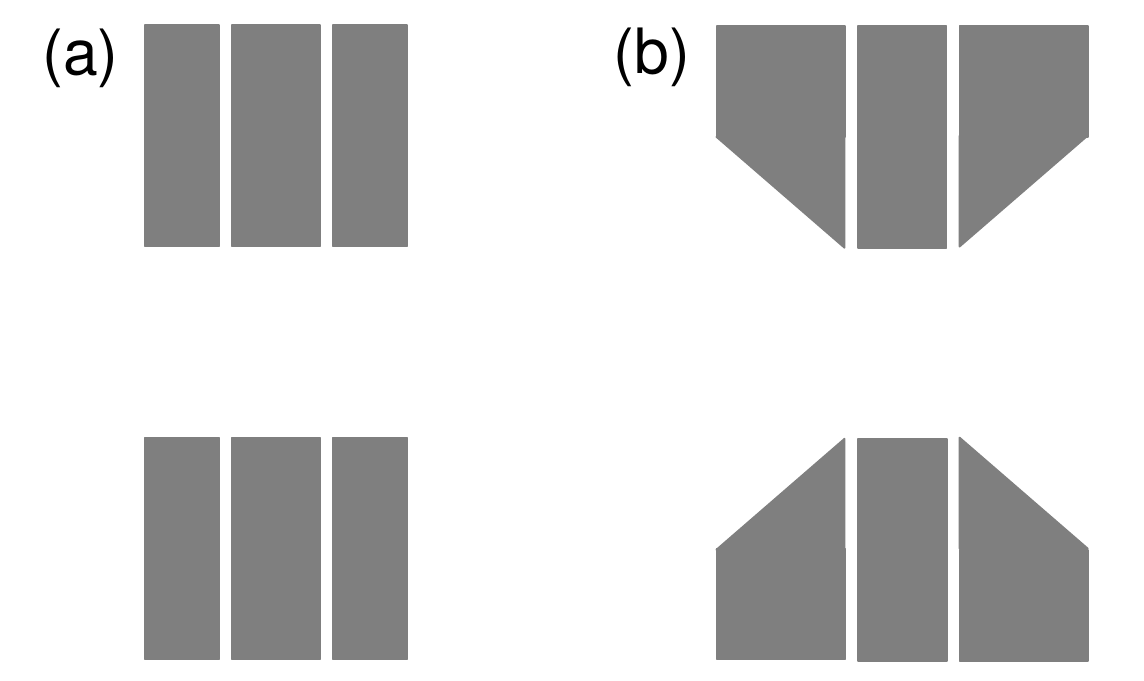}
\caption{(a) Design of the geometry of the 6 gate fingers for a
QPC$_{\rm 6F}$ device with 6 rectangular gate fingers. (b) Design of
a QPC$_{\rm 6F}$ device with the 4 outer gates in a shape that
explicitly induces a funnel shape for the entry and exit of the QPC
transport channel.} \label{fig2DesignQPC6F}
\end{figure}



\section{\label{5sec:designConsids}Design considerations}
We designed our QPC$_{\rm 6F}$ devices with 6 rectangular gate
fingers, in a symmetric layout with two sets of 3 parallel gate
fingers (Fig.~\ref{fig1SEMpic}b and Fig.~\ref{fig2DesignQPC6F}a).
SEM inspection of fabricated devices yields that the central gate
finger is 200~nm wide (as measured along the direction of channel
length $L_{eff}$). The outer gate fingers are 160~nm wide, and the
narrow gaps between gate fingers are 44~nm wide. This yields $(200 +
2 \cdot 160 + 2 \cdot 44)~{\rm nm} = 608~{\rm nm}$ for the total
distance between the outer sides of the 3 parallel gate fingers. The
lithographic width of the QPC channel (distance between the two sets
of 3 gate fingers) is 350~nm.

An example of alternative designs for the gate geometries that we
considered  is in Fig.~\ref{fig2DesignQPC6F}b. This design has a
two-sided funnel shape for the channel and this could result in
length-tunable QPC operation that better maintains a regular shape
for the saddle-point potential. However, the electrostatic
simulations in Section~\ref{5sec:Simulations} show that the
rectangular gate fingers as in Fig.~\ref{fig2DesignQPC6F}a also give
a length-tunable saddle-point potential that maintains a regular
shape while tuning the length. This observation holds for a range of
device dimensions similar to our design. For our particular design,
the lithographic length and width (350~nm) of the channel are
comparable, and the 2DEG is as far as 110~nm distance below the
surface (and the part in the center of the channel that actually
contains electrons is very narrow, about 20~nm). In this regime, the
saddle-point potential is strongly rounded with respect to the
lithographic shapes of the gates (see for example
Fig.~\ref{fig3SaddlePotential}c,d). An important advantage of the
rectangular design is that it provides two clear points for
calibrating the effective channel length $L_{eff}$: Operating at
$V_{g2}/V_{g1} = 0$ gives $L_{eff}=L_{litho}$ for the central gate
finger alone (200~nm, see Fig.~\ref{fig1SEMpic}b), while operating
at $V_{g2}/V_{g1} = 1$ gives $L_{eff}$ equal to the lithographic
distance between the outer sides of the 3 parallel gate fingers
(608~nm).

A point of concern for this design that deserves attention is
whether the narrow gaps between the 3 parallel gate fingers induce
significant structure on the saddle-point potential. The
electrostatic simulations show that this is not the case (see again
the examples in Fig.~\ref{fig3SaddlePotential}c,d). The part of the
channel that contains electrons is relatively far away from the gate
electrodes, and the potential $U$ at this location is strongly
rounded. Notably, the full height of the potentials in
Fig.~\ref{fig3SaddlePotential} is about 1~eV, while the occupied
subbands are at a height of only about 10~meV above the stationary
point of the saddle-point potential (in the center of the channel).
Such gaps between parallel gate electrodes can be much narrower when
depositing a wider gate on top of the central gate, with an
insulating layer between them. We chose against applying this idea
since we also aimed to have devices with a very low level of noise
and instabilities from charge fluctuations at defect and impurity
sites in the device materials. In this respect, we expect better
behavior when all gate fingers are deposited in a single fabrication
cycle, and when deposition of an insulating oxide or polymer layer
can be omitted.


\section{\label{5sec:Simulations}Electrostatic simulations}

This section presents results of electrostatic simulations of the
saddle-point potentials that define the QPC channel. The focus is on
the design with 6 rectangular gate fingers
(Fig.~\ref{fig2DesignQPC6F}a), with gate dimensions as mentioned in
the beginning of Section~\ref{5sec:designConsids}. The simulations
are based on the modeling approach that was introduced by Davies
\textit{et al.} \cite{5Davies}.

\subsection{Davies' method for simulating 2DEG electrostatics}

Davies \textit{et al.} \cite{5Davies} introduced a method for
modeling the electrostatics of gated 2DEG. It calculates the
electrostatic potential $U$ for electrons in the 2DEG regions around
the gates (the approach only applies to the situation where the 2DEG
underneath the gates is depleted due to a negative voltage on gate
electrodes). There are other models and approaches
\cite{5JohnAltDavies,5SniderAltDavies,5MinhanAltDavies,5LauxAltDavies,5John1AlteDavies}
for calculating such potential landscapes, but these are all more
complicated and computationally more demanding. The approach by
Davies \textit{et al.} is relatively simple. It does not account for
electrostatic screening effects, and, notably, it does not account
for the electron many-body interactions that were mentioned earlier.
Still, it was shown that it is well suited for calculating a valid
picture of a QPC saddle-point potential near the channel pinch-off
situation \cite{5Koop}.


\begin{figure}[t!]
\centering
\includegraphics[width=0.85\columnwidth]{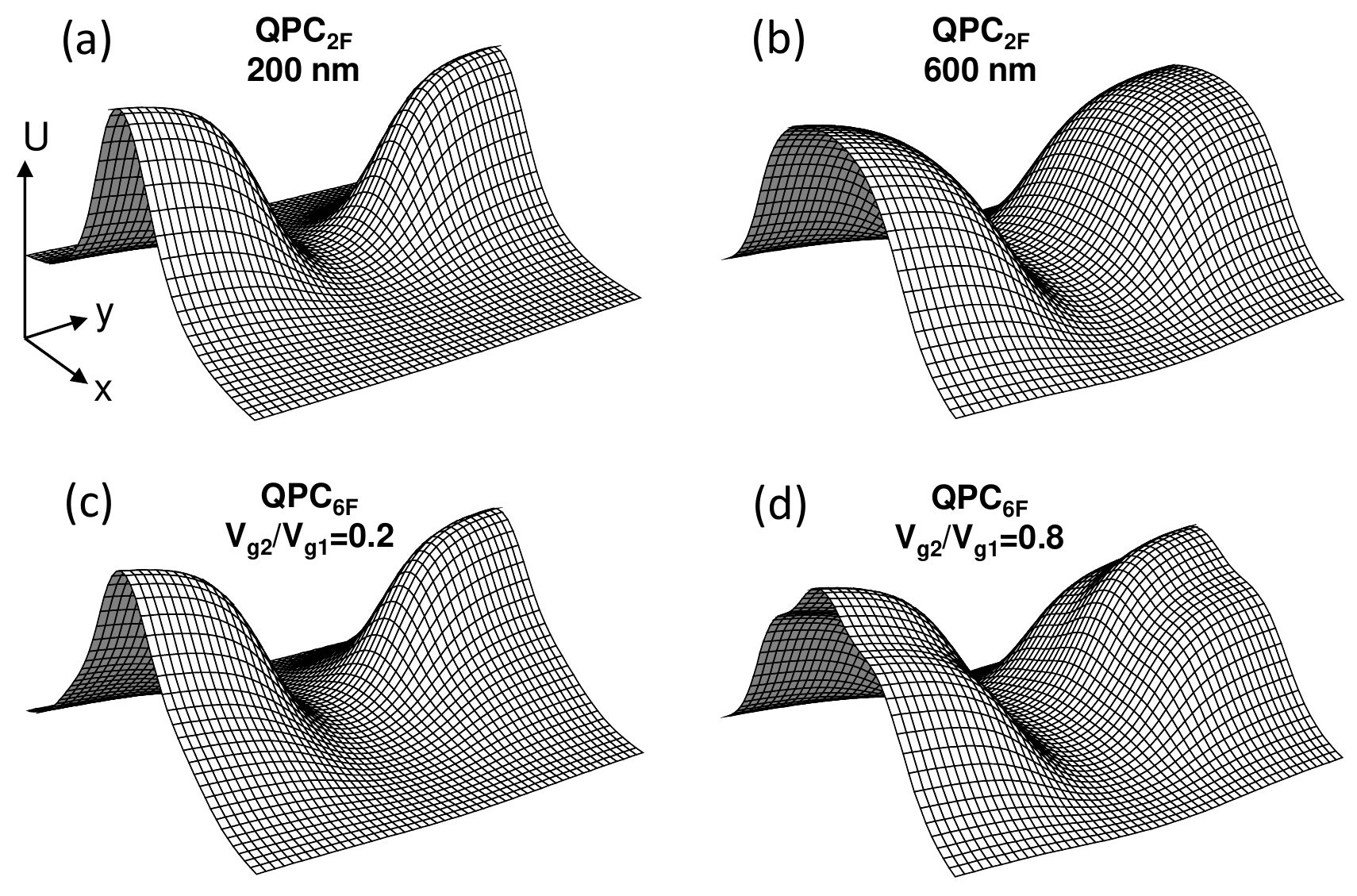}
\caption{(a),(b) Saddle-point potentials that represent the
electrostatic potential $U$ felt by electrons in the 2DEG plane. The
plots represent an area of 1000$\times$1000~${\rm nm^2}$, centered
at the middle of a QPC channel with a length $L_{litho}$ of 200~nm
(a) and 600~nm (b) of a QPC$_{\rm 2F}$ device with a lithographic
channel width of 350~nm. It is calculated for the material
parameters that are valid for the measured devices. See
Fig.~\ref{fig1SEMpic}b for relating the $x$- and $y$-direction to
the gate geometry. (c),(d) Similar saddle-point potentials $U$
calculated for QPC$_{\rm 6F}$ devices (with material parameters and
geometry as the measured devices). The effective channel length is
shorter for the case that is calculated for $V_{g2}/V_{g1}=0.2$ (c)
than for the case $V_{g2}/V_{g1}=0.8$ (d) (also note that QPC$_{\rm
6F}$ results for $V_{g2}/V_{g1}=0$ are the same as plot (a)). Panel
(c) and (d) also show that the narrow gaps between 3 parallel gate
fingers do not induce significant structure at low energies in the
saddle-point passage (it only induces a weak fingerprint off to the
side in the channel, at energies that are much higher than the
occupied electron levels, see panel (d)).}
\label{fig3SaddlePotential}
\end{figure}


The negative voltage on a gate that is needed to exactly deplete
2DEG underneath a large gate is called the threshold voltage
$V_{t}$, and it is to a good approximation given by
\begin{equation}
V_{t} = \frac{-e n_{2D}d}{\epsilon_r\epsilon_0}. \label{5equ:Vt}
\end{equation}
Here $n_{2D}$ is the electron density in the 2DEG (at zero gate
voltage), $d$ is the depth of the 2DEG, $\epsilon_r$ is the relative
dielectric constant of the material below the gate, and $\epsilon_0$
is the dielectric constant of vacuum (for details see
Ref.~\onlinecite{5Koop,5Davies}). The value of $V_t$ for a certain
2DEG material defines the value $U_0$ where the electrostatic
potential $U$ for electrons in the 2DEG becomes higher than the
chemical potential of the 2DEG. In turn, this can be used to define
in an arbitrary potential landscape $U$ (for arbitrary gates shapes
and for arbitrary gate voltages) the positions where $U=U_0$. That
is, one can calculate the positions in a gated device structure
where there is a boundary between depleted and non-depleted 2DEG,
and also calculate the electrostatic potential $U$ around such
points. When the center of the QPC has $U=U_0$, the channel is at
pinch-off and no electrons can pass through the QPC. The gate
voltage at which this happens is called the pinch-off voltage
$V_{po}$. Notably, the calculated value of $U$ at a certain position
is simply the superposition of all the contributions to $U$ from
different gate electrodes, and it is linear in the gate voltage on
each of these electrodes \cite{5Davies}.

Figure~\ref{fig3SaddlePotential} presents examples of saddle-point
potentials $U$ that are calculated with Davies' method, both for
QPC$_{\rm 2F}$ and QPC$_{\rm 6F}$ devices. The calculations are for
material parameters and geometries of measured devices (as described
in detail in the next sections).
Figures~\ref{fig3SaddlePotential}c,d show that the length of the
transport channel depends on the applied ratio $V_{g2}/V_{g1}$, and
that the narrow gaps between 3 parallel electrodes in QPC$_{\rm 6F}$
devices do not give significant structure on the saddle-point
potential in the operation regime that we consider.

\subsection{Definition and tuning of the effective length $L_{eff}$}

The focus of this work is on realizing QPC channels with a tunable
length. The channels are in fact saddle-point potentials (see
Fig.~\ref{fig3SaddlePotential}), and it is for such a smooth shape
not obvious what the value is of the channel length. We therefore
characterize this channel length with the parameter $L_{eff}$, which
corresponds to the value of the lithographic length $L_{litho}$ of a
QPC$_{\rm 2F}$ type device (with rectangular gate electrodes, see
Fig.~\ref{fig1SEMpic}a) that gives effectively the same saddle-point
potential.


\begin{figure}[!]
\centering
\includegraphics[width=0.85\columnwidth]{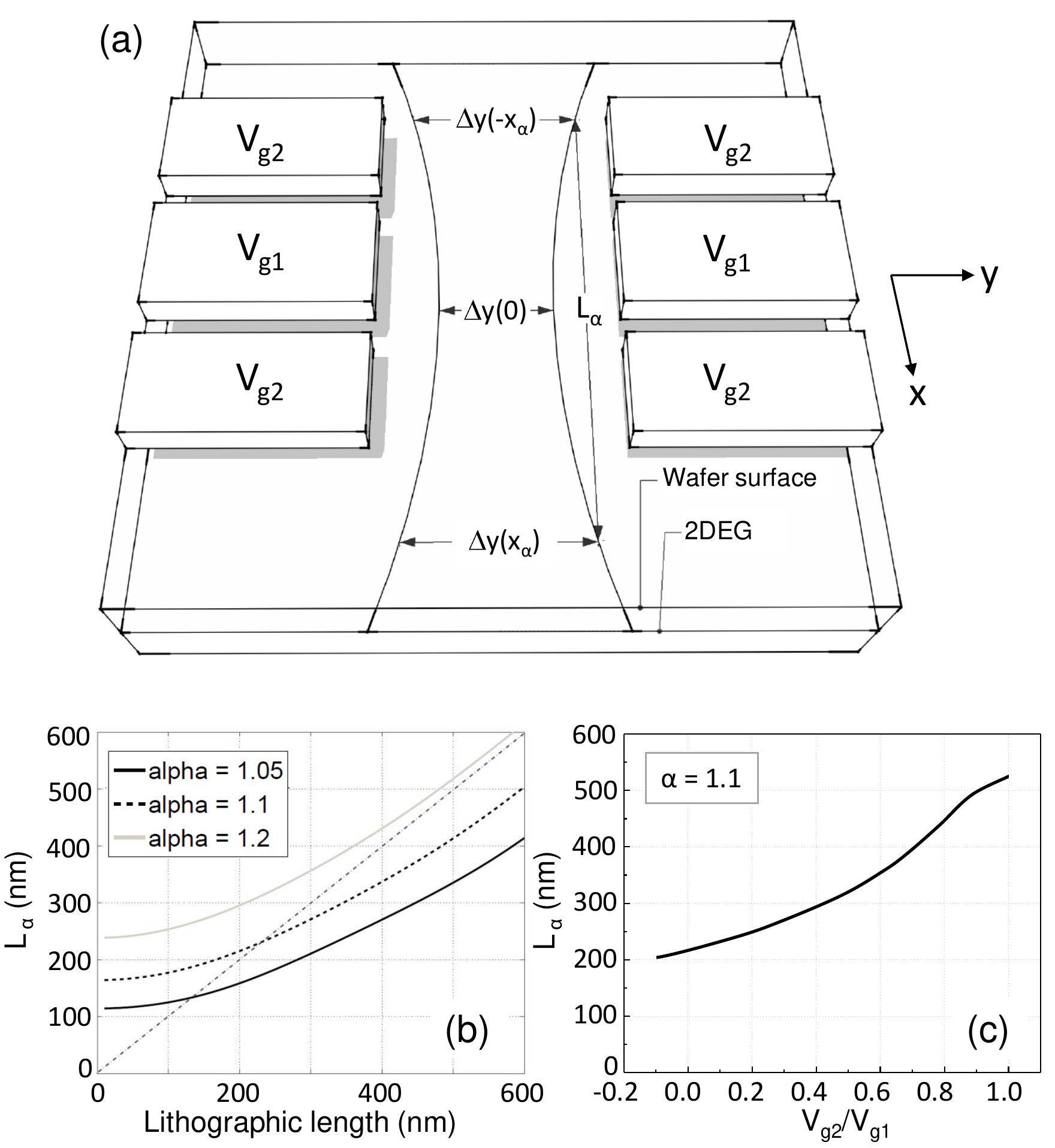}
\caption{(a) Schematic representation of a QPC$_{\rm 6F}$ device,
illustrating length variables that are introduced in the main text.
(b) The calculated length $L_{\alpha}$ for a range of values of the
lithographic length of QPC$_{\rm 2F}$ devices, for three values of
$\alpha$. (c) The calculated effective length $L_{\alpha}$ for
$\alpha=1.1$ for a QPC$_{\rm 6F}$ device, as a function of the ratio
$V_{g2}/V_{g1}$.} \label{5fig:Lalpha}
\end{figure}


We implemented this as follows. We calculated the saddle-point
potential $U(x,y)$ for the pinch-off situation (see
Fig.~\ref{fig1SEMpic}b and Fig.~\ref{fig3SaddlePotential}a for how
the $x$- and $y$-directions are defined). The transverse confinement
in the middle of the QPC (defined as $x=0$, $y=0$) is parabolic to a
very good approximation. When moving out of the channel along the
$x$-direction, the transverse confinement becomes weaker, but
remains at first parabolic. Notably, the energy eigenstates for
confinement in such a parabolic potential, described as
\begin{equation}
U(y)= \frac12 m^* \omega_0^2 y^2,
\label{5equ:parabollaUy}
\end{equation}
have a width that is (for all levels) proportional to
$\omega_0^{-1/2}$. In this expression $m^*$ is the effective mass of
the electron and $\omega_0$ is the angular frequency of natural
oscillations in this potential. The parameter $\omega_0$ defines
here the steepness of $U(y)$, and we obtain $\omega_0 (x)$ values
from fitting Eq.~\ref{5equ:parabollaUy} to potentials $U(x,y)$
obtained with Davies's method. We use this and investigate the width
$\Delta y(x)$ in $y$-direction for the lowest energy eigenstate, at
all positions $x$ along the channel (see Fig.~\ref{5fig:Lalpha}a).
For parabolic confinement this wavefunction in $y$-direction has a
Gaussian shape and has a width
\begin{equation}
\Delta y(x) = \sqrt{\frac{h}{4 \pi \; m^* \; \omega_0(x)}}.
\label{5equ:widthDeltaY}
\end{equation}

With this approach we analyzed that the distance from $x=0$ to the
$x$-position $x_{\alpha}$ where the value $\Delta y(x)$ increased by
a factor $\alpha \approx 1.1$ defines a suitable point for defining
the value of $L_{eff}$. That is, we define
\begin{equation}
L_{\alpha} = 2 \; x_{\alpha},
\label{5equ:DEFLalpha}
\end{equation}
and find $x_{\alpha}$ by solving
\begin{equation}
\Delta y(x=x_{\alpha}) = \alpha \cdot \Delta y(x=0)
\label{5equ:SolveXalpha}
\end{equation}
for a certain $\alpha$. Subsequently, $L_{eff}$ is defined by using the suitable $\alpha$ value,
\begin{equation}
L_{eff} = L_{\alpha} \; \;{\rm for} \; \; \alpha=1.1.
\label{5equ:DEFLeffalpha}
\end{equation}
We came to this parameterization as follows. We used this
\textit{ansatz} first in simulations of QPC$_{\rm 2F}$ devices.
Here, we explored for different values of $\alpha$ the relation
between $L_{litho}$ and $L_{\alpha}$. Results of this for $\alpha =
1.05$, 1.1 and 1.2 are presented in Fig.~\ref{5fig:Lalpha}b. For the
range of $L_{litho}$ values that is of interest to our study
($\sim$100~nm to $\sim$500~nm), we find the most reasonable overall
agreement between the actual value for $L_{litho}$ (input to the
simulation) and the value $L_{\alpha}$ (derived from the simulation)
for $\alpha=1.1$. The agreement is not perfect, but we analyzed that
the deviation is within an uncertainty that we need to assume
because the exact shapes of saddle-point potentials in different
device geometries do show some variation, and because the limited
validity of Davies' method. Nevertheless, it provides a reasonable
recipe for assigning a value $L_{eff}$ to any saddle-point
potential, with at most 20\% error.

Fig.~\ref{5fig:Lalpha}c presents results of calculating
$L_{\alpha}=L_{eff}$ for $\alpha=1.1$ from simulations of a
QPC$_{\rm 6F}$ device, operated at different values for
$V_{g2}/V_{g1}$. The results show a clear monotonic trend, with
$L_{eff}=210~{\rm nm}$ for $V_{g2}/V_{g1}=0$ to $L_{eff}=525~{\rm
nm}$ for $V_{g2}/V_{g1}=1$. This is for a QPC$_{\rm 6F}$ device for
which we expect $L_{eff}=200~{\rm nm}$ for $V_{g2}/V_{g1}=0$ and
$L_{eff}=608~{\rm nm}$ for $V_{g2}/V_{g1}=1$ (see
Section~\ref{5sec:designConsids}). In
Section~\ref{5sec:ExperimentalRelization} we discuss how this latter
point is used for applying a small correction to the simulated
values for $L_{eff}$. These simulations show that the QPC$_{\rm 6F}$
that we consider allows for tuning $L_{eff}$ by about a factor 3.

It is worthwhile to note that our current design showed optimal
behavior in the sense that it can tune $L_{eff}$ from about 200~nm
to 600~nm, while the dependence of $L_{eff}$ on $V_{g2}/V_{g1}$ is
close to linear. We also simulated QPC$_{\rm 6F}$ devices with wider
gate electrodes for the outer gates, and (as mentioned in
Section~\ref{5sec:designConsids}) devices with gate geometries as in
Fig.~\ref{fig2DesignQPC6F}b. These devices showed a steeper slope
for part of the relation between $V_{g2}/V_{g1}$ and $L_{eff}$,
which is not desirable.

\section{\label{5sec:sampleFabAndTech}Sample fabrication and measurement techniques}

We fabricated QPC devices with a GaAs/${\rm Al}_{0.35}{\rm
Ga}_{0.65}{\rm As}$ MBE-grown heterostructure, which has a 2DEG at
110~nm depth below its surface from modulation doping. The layer
sequence and thickness of the materials from top to bottom
(\textit{i.e.} going into the material) starts with a $5$ nm GaAs
capping layer, then a 60~nm ${\rm Al}_{0.35}{\rm Ga}_{0.65}{\rm As}$
layer with Si doping at about $1 \times 10^{18} \text{cm}^{-3}$,
which is followed by an undoped spacer layer of 45~nm. The 2DEG is
located in a heterojunction quantum well at the interface with the
next layer, which is a 650~nm undoped GaAs layer. This
heterostructure was grown on a commercial semi-insulating GaAs
wafer, after first growing a sequence of 10 GaAs/AlAs layers for
smoothing the surface and trapping impurities. The 2DEG had an
electron density $n_{2D}=1.6 \cdot 10^{15}~{\rm m^{-2}}$ and a
mobility $\mu = 118~{\rm m^2 V^{-1} s^{-1}}$. We fabricated both
conventional QPC$_{\rm 2F}$ devices and QPC$_{\rm 6F}$ devices by
standard electron-beam lithography and clean-room techniques. The
gate fingers were deposited using 15~nm Au on top of a 5~nm Ti
sticking layer. For measuring transport through the QPCs we realized
ohmic contacts to the 2DEG reservoirs by annealing of a AuGe/Ni/Au
stack that was deposited on the wafer surface
\cite{5OhmicKoopIqbal}. The geometries of the fabricated devices
were already described in the beginning of
Section~\ref{5sec:designConsids}.

The measurements were performed in a He-bath cryostat and in a
dilution refrigerator, thus getting access to effective electron
temperatures from 80~mK to 4.2~K. We used standard lock-in
techniques with an a.c.~excitation voltage $V_{bias}=10~{\rm \mu V}$
RMS at $387$~Hz. Fig.~\ref{fig1SEMpic}a shows the 4-probe
voltage-biased measurement scheme, where both the current and the
actual voltage drop $V_{sd}$ across the QPC channel are measured
such that any influence of series resistances could be removed
unambiguously. The gate voltages are applied with respect to a
single grounded point in the loop that carries the QPC current.

\section{\label{5sec:ExperimentalRelization}Experimental realization of length-tunable QPCs}


\begin{figure}[t!]
\centering
\includegraphics[width=0.85\columnwidth]{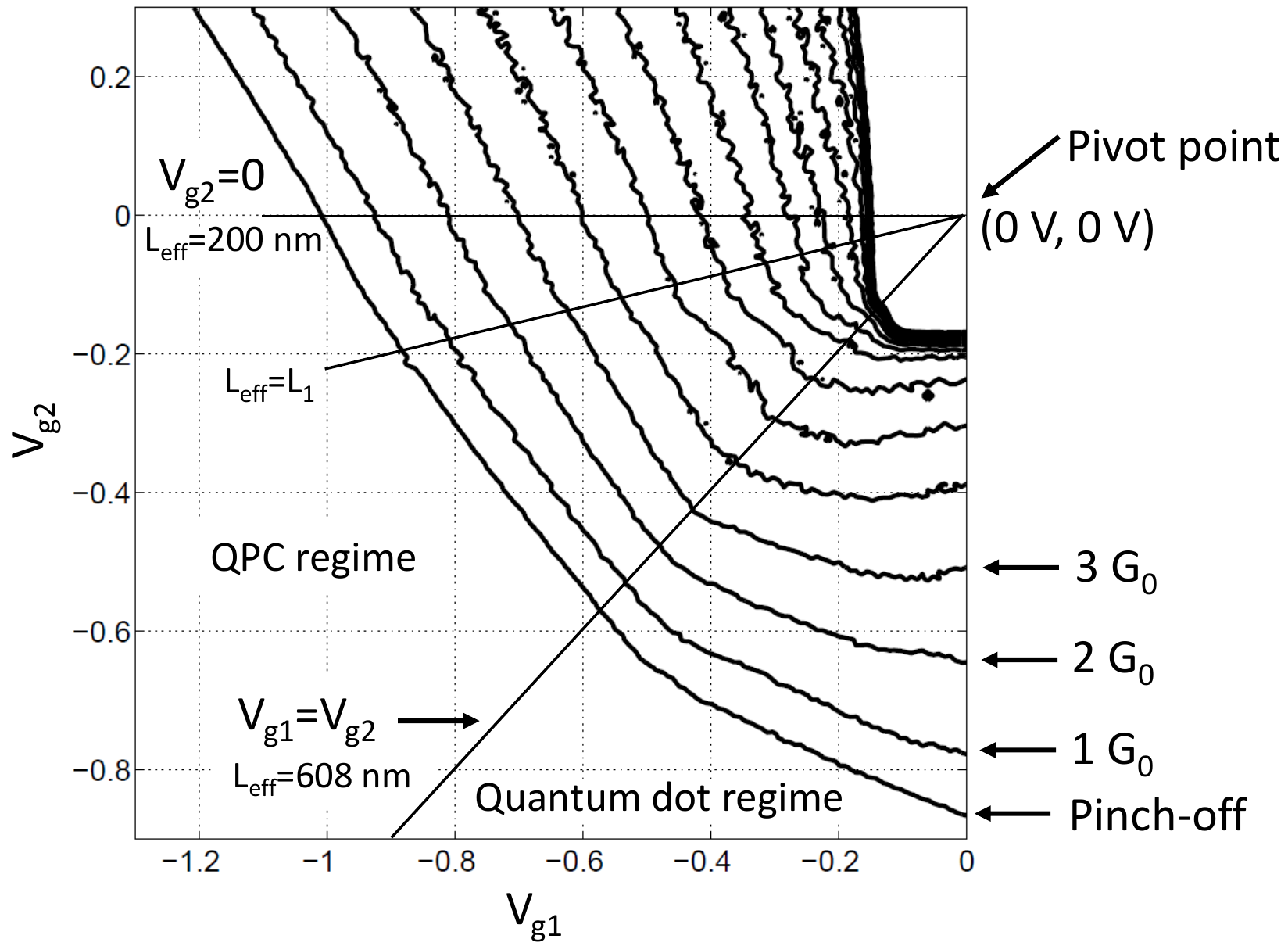}
\caption {QPC linear conductance as a function of $V_{g1}$ and
$V_{g2}$ for a QPC$_{\rm 6F}$ device, presented in the form of
iso-conductance lines at integer $G_0$ levels. The conductance was
measured at 4.2~K where the quantized conductance is nearly fully
washed out by temperature. The two operational regimes above and
below the line $V_{g1}=V_{g2}$ yield QPC and quantum dot behavior,
respectively.} \label{5fig:Isoconductance2D}
\end{figure}

This section presents an experimental characterization of the
QPC$_{\rm 6F}$ devices that we designed (Fig~\ref{fig1SEMpic}b) and
we compare the results to our simulations.
Figure~\ref{5fig:Isoconductance2D} presents measurements of the
conductance $G$ as a function of $V_{g1}$ and $V_{g2}$. Several
labels in the plot illustrate relevant concepts, which were partly
discussed before. For the area in this plot with $V_{g2}$ more
negative than $V_{g1}$ we expect some quantum-dot like localization
in the middle of the channel and this regime should therefore be
avoided in studies of QPC behavior. Further, the plot illustrates
that operation for a particular value of $L_{eff}$ requires
co-sweeping of $V_{g1}$ and $V_{g2}$ from a particular point below
pinch-off in a straight line to the pivot point. This corresponds to
opening the QPC at a fixed ratio for $V_{g2}/V_{g1}$. The pivot
point is the point where the gate voltages do not alter the original
electron density of the 2DEG. For this measurement this is for
$V_{g1}=V_{g2}= 0~{\rm V}$, but this is different for the case of
biased cool downs. We carried out biased cool downs for suppressing
noise from charge instabilities in the donor layer
\cite{5PRBBCool,5PRLBCool}. For such experiments the QPCs were
cooled down with a positive voltage on the gates. We typically used
+0.3~V, and observed indeed better stability with respect to charge
noise. The effect of such a cool down can be described as a
contribution to the gate voltage of -0.3~V that is frozen into the
material \cite{5PRBBCool,5PRLBCool}. Consequently, co-sweeping of
$V_{g1}$ and $V_{g2}$ for maintaining a fixed channel length must
now be carried out with respect to the pivot point $V_{g1}=V_{g2}= +
0.3~{\rm V}$ instead of $V_{g1}=V_{g2}= 0~{\rm V}$.


\begin{figure}[t!]
\centering
\includegraphics[width=0.85\columnwidth]{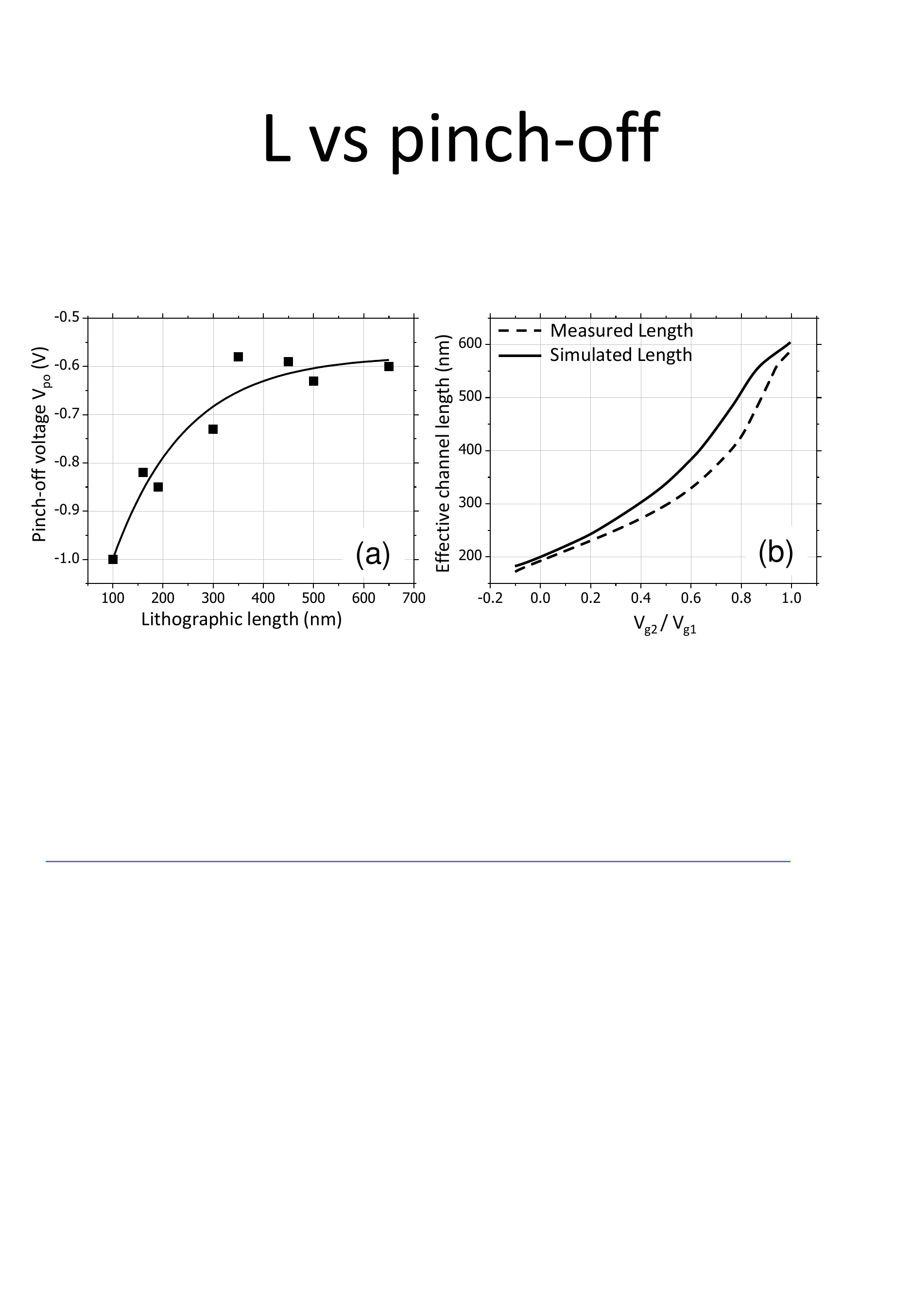}
\caption{(a) Experimentally determined relation between the
pinch-off gate voltage $V_{po}$ and the lithographic length of
QPC$_{\rm 2F}$ devices. Points are experimental results. The solid
line is a phenomenological expression that was used for
parameterizing the relation between $V_{po}$ and the lithographic
length. (b) Comparison between measured and simulated values of the
effective channel length for a QPC$_{\rm 6F}$ device.}
\label{5fig:LalphaVsmeasured}
\end{figure}


The theory behind the Davies method illustrates why operation at
fixed effective lenght requires a fixed ratio $V_{g2}/V_{g1}$. All
points in the potentials landscapes $U$ for QPC$_{\rm 2F}$ devices
as in Fig.~\ref{fig3SaddlePotential}a,b have a height that scales
linear with the gate voltage $V_g$. Thus, when opening the QPC the
full saddle-point potential changes height at a fixed shape.
Mimicking this situation with QPC$_{\rm 6F}$ devices requires a
fixed ratio $V_{g2}/V_{g1}$, again because $V_{g1}$ and $V_{g2}$
influence $U$ in a linear manner. The plot also illustrates the two
special operation lines where the effective length of the channel is
unambiguous, and we used these points to better calibrate the
relation between $V_{g2}/V_{g1}$ and $L_{eff}$. The first case is
the line at $V_{g2}=0$, which yields $L_{eff}=200~{\rm nm}$, as
defined by the central gates alone. The second case is the line
$V_{g1}=V_{g2}$. Here $L_{eff}$ is 608~nm, as defined by the full
lithographic length of the 3 gate fingers.

\begin{figure}[t!]
\centering
\includegraphics[width=1.00\columnwidth]{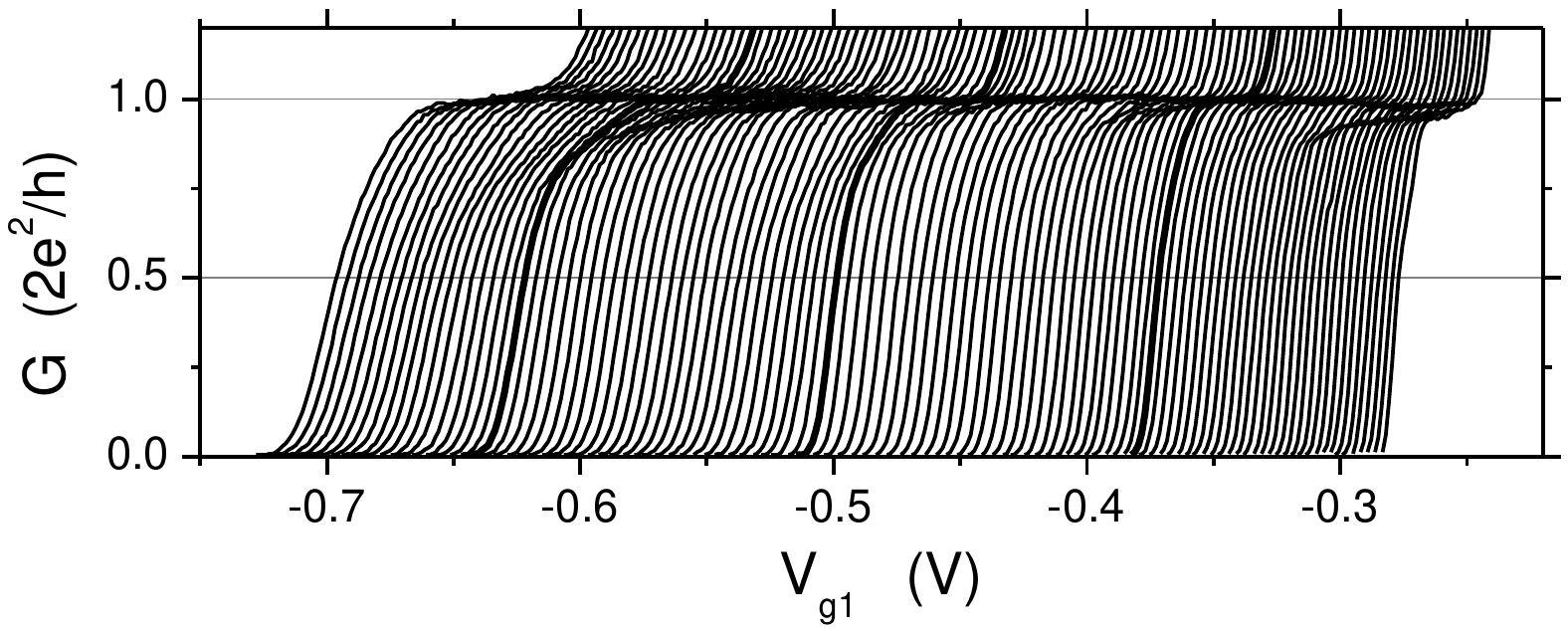}
\caption{The linear conductance $G$ as a function of $V_{g1}$ for a
QPC$_{\rm 6F}$ device, measured at 80~mK. Different traces are for
fixed ratio $V_{g2}/V_{g1}=0$ to $V_{g2}/V_{g1}=1$ (left to right,
traces \textit{not} offset), which corresponds to increasing the
effective channel length $L_{eff}$ from 200~nm to 608~nm.}
\label{5fig:LinearG}
\end{figure}

We improved and further checked our calibration of the relation
between $V_{g2}/V_{g1}$ and $L_{eff}$ as follows. We used the trend
that came out of the simulations (Fig.~\ref{5fig:Lalpha}c), but
pinned the curve at 200~nm for $V_{g2}/V_{g1}=0$ and at 608~nm for
$V_{g2}/V_{g1}=1$ (black line in Fig.~\ref{5fig:LalphaVsmeasured}b).
This trace shows good agreement with results from an independent
check (dashed line) that used the pinch-off gate voltage $V_{po}$ as
an identifier for the effective length. This independent check used
data from a set of QPC$_{\rm 2F}$ devices for calibrating the
relation between $L_{litho}$ and $V_{po}$
(Fig.~\ref{5fig:LalphaVsmeasured}a). This shows the trend that
shorter QPC$_{\rm 2F}$ devices require a more negative gate voltage
to reach pinch-off \cite{5VpvsL}. We related this to the pinch-off
values in QPC$_{\rm 6F}$ devices. In particular, we analyzed the
pinch-off points on the $V_{g1}$ axis, and its dependence on
$V_{g2}/V_{g1}$ (see also Fig.~\ref{5fig:LinearG}). The results of
using this for assigning a certain $L_{eff}$ to each $V_{g2}/V_{g1}$
is the dashed line in Fig.~\ref{5fig:LalphaVsmeasured}b, and shows
good agreement with the values that were obtained from simulations.
We can thus assign a value to $L_{eff}$ for each $V_{g2}/V_{g1}$
with an absolute error that is at most 50~nm. Notably, the relative
error when describing the increase in $L_{eff}$ upon increasing
$V_{g2}/V_{g1}$ is much smaller.

The results in Fig.~\ref{5fig:LinearG} provide an example of linear
conductance measurements on a QPC$_{\rm 6F}$ device at 80~mK. The
traces show clear quantized conductance plateaus for all settings of
$L_{eff}$. Several of these linear conductance traces also show the
0.7~anomaly, and the strength of its (here rather weak) expression
shows a modulation as a function of $L_{eff}$ over about 3 periods.
A detailed study of this type of periodicity can be found in
Ref.~\onlinecite{5IqbalPerio}. This example illustrates the validity
and importance of our type of QPCs in studies of length-dependent
transport properties.


\section{\label{5sec:Conclusion}Conclusions}

We have developed and characterized length-tunable QPCs that are
based on a symmetric split-gate geometry with 6 gate fingers. Gate
structures with different shapes and dimensions can be designed
depending upon the required range for length tuning and for
optimizing the tuning curve. For our purpose (QPCs with an effective
channel length between about 200~nm and 600~nm, and 350~nm channel
width) we found that simple rectangular gate fingers are an
attractive choice. Our simulations and experimental results are in
close agreement. We were able to tune the effective length by about
a factor 3, from 200~nm to 608~nm. QPCs are the simplest devices
that show clear signatures of many-body physics, as for example the
$0.7$ anomaly and the Zero-Bias Anomaly (ZBA) \cite{Cronenwett2002}.
Our length-tunable QPCs provide an interesting platform for
systematically investigating these many-body effects. In particular,
these QPCs provide a method for studying the influence of the QPC
geometry without suffering from device-to-device fluctuations that
hamper such studies in conventional QPCs with 2 gate fingers.
Studies in this direction are presented in
Ref.~\onlinecite{5IqbalPerio}.

\section*{ACKNOWLEDGMENTS}
We thank Y.~Meir for discussions, B.~H.~J.~Wolfs for technical
assistance, and the German programs DFG-SPP 1285, Research school
Ruhr-Universit\"{a}t Bochum and BMBF QuaHL-Rep 01BQ1035 for
financial support. MJI acknowledges a scholarship from the Higher
Education Commission of Pakistan.


\bibliographystyle{apsrev}


\end{document}